\documentclass[letterpaper]{IEEEtran}
\usepackage{graphicx}
\usepackage{xcolor}

\usepackage{amsmath}
\usepackage{amssymb}
\usepackage{cite}
\usepackage[colorlinks=Flase]{hyperref}
\usepackage{bbm}
\usepackage{algorithm} 
\usepackage{algpseudocode}
\usepackage{caption}
\usepackage{steinmetz}
\usepackage{subcaption}
\usepackage{amsthm}
\usepackage[english]{babel}

\newcounter{relctr} 
\everydisplay\expandafter{\the\everydisplay\setcounter{relctr}{0}} 
\DeclareMathOperator*{\minimize}{minimize} 
\DeclareMathOperator*{\maximize}{maximize} 
\usepackage[figurename=Fig.]{caption}

\newcommand\labelrel[2]{%
  \begingroup
    \refstepcounter{relctr}%
    \stackrel{\textnormal{(\alph{relctr})}}{\mathstrut{#1}}%
    \originallabel{#2}%
  \endgroup
}
\AtBeginDocument{\let\originallabel\label} 

\ifCLASSINFOpdf
\else
\fi

\begin{document}

\title{Energy Beamforming for RF Wireless Power Transfer with Dynamic Metasurface Antennas}

\author{Amirhossein~Azarbahram,~\IEEEmembership{Student~Member,~IEEE,}
        Onel~L.~A.~López,~\IEEEmembership{Member,~IEEE}
       Richard~D.~Souza,~\IEEEmembership{Senior~Member,~IEEE,}
       Rui~Zhang,~\IEEEmembership{Fellow,~IEEE,}
        and~Matti~Latva-Aho,~\IEEEmembership{Senior Member,~IEEE}
\thanks{A. Azarbahram, O. L\'opez and M. Latva-Aho, Centre for Wireless Communications (CWC), University of Oulu, Finland, \{amirhossein.azarbahram, onel.alcarazlopez, matti.latva-aho\}@oulu.fi; R. D. Souza, Department of Electrical and Electronics Engineering, Federal University of Santa Catarina, Florianopolis, Brazil, richard.demo@ufsc.br; R. Zhang, School of Science and Engineering, Shenzhen Research Institute of Big Data, The Chinese University of Hong Kong, Shenzhen, Guangdong, China, rzhang@cuhk.edu.cn, and Department of Electrical and Computer Engineering, National University of Singapore, Singapore, elezhang@nus.edu.sg.}%
\thanks{This work is partially supported in Finland by the Finnish Foundation for Technology Promotion, Academy of Finland (Grants 348515 and 346208 (6G Flagship)), by the European Commission through the Horizon Europe/JU SNS project Hexa-X-II (Grant Agreement no. 101095759); and in Brazil by CNPq (402378/2021-0, 401730/2022-0, 305021/2021-4) and RNP/MCTIC 6G Mobile Communications Systems (01245.010604/2020-14).}}


\maketitle

\begin{abstract}

Radio frequency (RF) wireless power transfer (WPT) is a promising technology for charging the Internet of Things. Practical RF-WPT systems usually require energy beamforming (EB), which can compensate for the severe propagation loss by directing beams toward the devices. The EB flexibility depends on the transmitter architecture, existing a trade-off between cost/complexity and degrees of freedom. Thus, simpler architectures such as dynamic metasurface antennas (DMAs) are gaining attention. Herein, we consider an RF-WPT system with a transmit DMA for meeting the energy harvesting requirements of multiple devices and formulate an optimization problem for the minimum-power design. First, we provide a mathematical model to capture the frequency-dependant signal propagation effect in the DMA architecture. Next, we propose a solution based on semi-definite programming and alternating optimization. Results show that a DMA-based structure can outperform a fully-digital implementation and that the required transmit power decreases with the antenna array size, while it increases and remains almost constant with frequency in DMA and FD, respectively.

\end{abstract}


\IEEEpeerreviewmaketitle

\section{Introduction}

\IEEEPARstart{F}{uture} wireless systems must support uninterrupted operation of massive Internet of Things networks, providing reliable energy supply for low-power wireless devices. Radio frequency (RF) wireless power transfer (WPT) is a promising solution to achieve this goal. The end-to-end power transfer efficiency of an RF-WPT system is inherently low, calling for efficient techniques such as waveform optimization, energy beamforming (EB), and distributed antenna systems \cite{intro3, zhangclerckxfundam}.

EB can focus energy beams toward one or more energy harvesting (EH) devices at the same time. The EB flexibility and potential gains are determined by the transmitter architecture. Although a fully-digital (FD) structure provides the highest flexibility, {it requires a large number of RF chains, including power amplifiers (PAs), resulting in high complexity, cost, and power consumption}. 
The analog architecture alternatives incur much lower power consumption and cost {by reducing the number of RF chains, and utilizing only analog circuits as phase shifters}. However, they may not offer sufficient flexibility, thus, hybrid architectures have been introduced to combine digital and analog approaches and provide beamforming capability with a trade-off between complexity/cost and flexibility \cite{hybridbeamsurvey}. 

{In hybrid architectures, RF chains and phase shifters can be connected using, e.g., the fully-connected network, where all RF chains are connected to all phase shifters. However, since the complexity and losses of analog circuits affect scalability and performance}, simpler array of subarrays (AoSA)-based architectures are used~\cite{arraysubarray1}. Another technology that avoids analog circuits is the dynamic metasurface antenna (DMA), a group of configurable metamaterial elements placed on a set of waveguides \cite{DMAbase}. {Unlike reflective surfaces  \cite{IRS_Rui}, where there is no correlation between the phase shift and gain introduced by each reflecting element, the modeling of DMA metamaterial elements is more involved, making their corresponding design problems different in general.} DMA is utilized in \cite{DMAWPT} in near-field WPT to maximize the sum-harvested energy considering linear {RF-to-direct current (DC)} 
power conversion. However, in practice, there are extensive non-linearities in the EH device. Thus, a linear power conversion model cannot represent the amount of harvested power in a practical EH device. Herein, we also consider a DMA transmitter architecture to charge multiple single-antenna EH devices but utilize a practical alternative, which assumes that each device has a specific DC EH requirement and that the corresponding RF power requirement can be obtained by the nonlinear EH relationship of the device. Then, providing that specific RF power required by the device leads to meeting the EH requirements.

Our main contributions are: i) we optimize the EB aiming to minimize the transmit power while meeting the EH requirements of all receivers; ii) we provide a frequency-dependant model for the propagation characteristics of the DMA, which was previously overlooked; iii) we propose an efficient EB solution relying on semi-definite programming (SDP) and alternating optimization, which converges in polynomial time; iv) we show that DMA outperforms the FD structure in terms of minimum transmit power for meeting the RF power requirements, and that the transmit power reduces with increasing the antenna array size, while it increases with the operating frequency in the DMA architecture. Next, Section~\ref{sec:SystemModel} presents the system model and the optimization problem. In Section~\ref{sec:optframework}, we discuss the proposed solution, while Section~\ref{sec:results} provides numerical results. Finally, Section~\ref{sec:consclusion} concludes the paper.

\textbf{Notations}: Bold lowercase (upper-case) letters represent vectors (matrices), $| \cdot |$ is the $l_2$ norm, $(\cdot)^T$ is the transpose and $(\cdot)^H$ is the transpose conjugate. $\mathbb{E}$ is the expectation, $\otimes$ is the Kronecker product, $\mathrm{rank}(\cdot)$ and $\mathrm{Tr}(\cdot)$ are the rank and trace of a matrix, respectively. $\mathrm{Vec}(\cdot)$ is the vectorization operation.

\vspace{-3mm}
\section{System Model \& Problem Formulation}\label{sec:SystemModel}
\vspace{-1mm}

\begin{figure}[t]
    \centering
    \includegraphics[width=0.9\columnwidth]{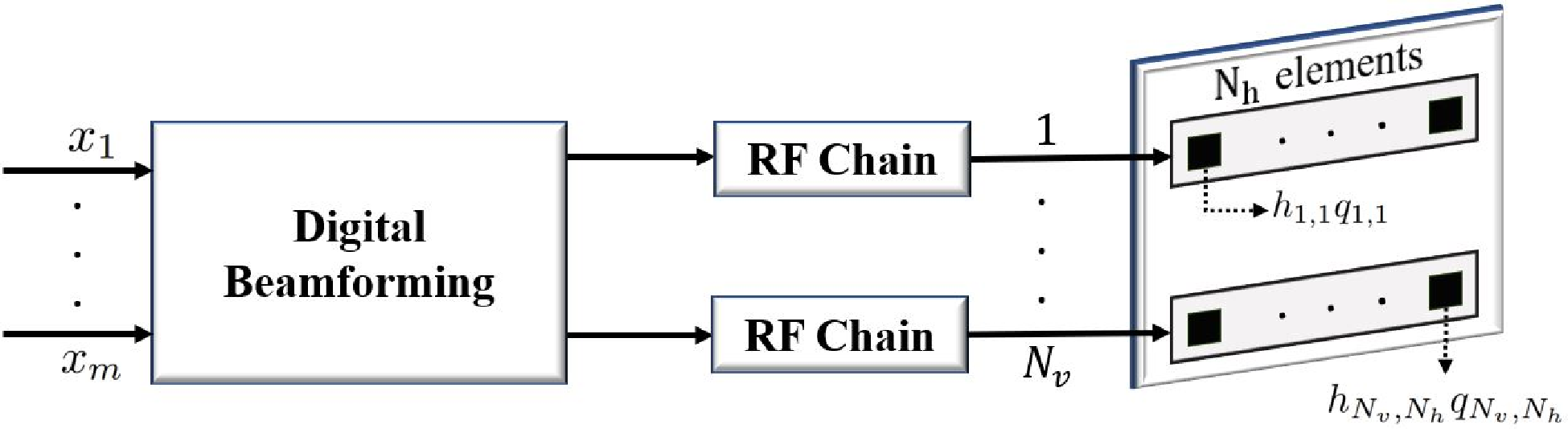}
    \caption{DMA-based transmitter architecture.}
    \label{fig:TXArch}
    \vspace{-5mm}
\end{figure}

We consider a multi-antenna RF-WPT system where a transmitter with a DMA uniform square array charges $K$ single-antenna devices. As illustrated in Fig.~\ref{fig:TXArch}, $M$ energy symbols are the inputs of a digital beamformer, while there are $N_v$ RF chains at the output. Each RF~chain is connected to a waveguide with $N_h$ metamaterial elements. Thus, the total number of elements is $N = N_v \times N_h$.

\subsubsection{Channel Model}\label{subsec:chan}

We consider the near-field wireless channel. The $k$-th user, located at a distance $r_k$ from the transmitter, lies in the near-field propagation region if $d_{fs} < r_k < d_{fr}$, where $d_{fs} = \sqrt[3]{\frac{D^4}{8\lambda}}$ and $d_{fr} = \frac{2D^2}{\lambda}$ are the Fresnel and Fraunhofer distance, respectively. Moreover, $\lambda$ is the wavelength, $D = \sqrt{2}L$ is the antenna diameter, and $L$ is the antenna length. The location of the $l$th element in the $i$th waveguide is $\mathbf{p}_{i,l} = [x_{i,l}, y_{i,l}, z_{i,l}]^T$, $i = 1,2,\ldots,N_v$ and $l = 1,2,\ldots, N_h$, while $\mathbf{p}_{k}$ denotes the $k$th device location.

The channel coefficient between user $k$ and the $l$th metamaterial element in the $i$th waveguide is given by \cite{near-field}
\begin{equation}\label{eq:channelcoef}
    \gamma_{i,l,k} = A_{i,l,k} e^{-j2\pi d_{i,l,k}/\lambda},
\end{equation}
where $2\pi d_{i,l,k}/\lambda$ constitutes the phase shift introduced by the propagation distance, and $d_{i,l,k} = |\mathbf{p}_k - \mathbf{p}_{i,l}|$ is the Euclidean distance between the element and the user. Additionally, the channel gain coefficient is  
\begin{equation}
    A_{i,l,k} = \sqrt{F(\theta_{i,l,k},\phi_{i,l,k})}\frac{\lambda}{4\pi d_{i,l,k}},
\end{equation}
where $\theta_{i,l,k}$ and $\phi_{i,l,k}$ are the elevation and azimuth angle, respectively, while $F(\theta_{i,l,k},\phi_{i,l,k})$ denotes the radiation profile of each element. We assume that the latter is given by \cite{anetnna_radiation}
\begin{equation}
        F(\theta_{i,l,k},\phi_{i,l,k}) = \begin{cases}
        G_t{\cos{(\theta_{i,l,k})}}^{\frac{G_t}{2}-1}, & \theta_{i,l,k} \in [0,\frac{\pi}{2}],
        \\
        0, & \text{otherwise},
        \end{cases}
\end{equation}
where $G_t = 2(b+1)$ and $b$ are the transmit antenna gain and the boresight gain, respectively. Now,  let us define $\boldsymbol{\gamma}_k = \begin{bmatrix}
\gamma_{1,1,k} ,\gamma_{1,2,k}, \ldots, \gamma_{N_v,N_h,k}\end{bmatrix}^T$ as the $N$-dimensional channel coefficients vector between the user $k$ and the elements of the DMA. Note that under the conventional far-field conditions, the channel coefficient can be represented as $A_{k} e^{-j\psi_{i,l,k}}$, where $A_k$ is only determined by the distance between the user and the antenna array, and $\psi_{i,l,k}$ by the spatial direction of the user and the spacing between the antenna elements.
\subsubsection{DMA Model}\label{subsec:DMAcharach}

In practice, microstrip lines are usually utilized as the DMA waveguides. The propagation coefficient  for the $l$th element in the $i$th microstrip is given by \cite{DMAformulation}
\begin{equation}\label{eq:propWG}
    h_{i, l} = e^{-(l-1)d_l(\alpha_i+j\beta_i)},
\end{equation}
where $d_l$, $\alpha_i$, and  $\beta_i$ are the inter-element distance, waveguide attenuation coefficient, and propagation constant, respectively. We assume that all microstrips are of the same type; $\beta_i = \beta$ and $\alpha_i = \alpha$. In general, the frequency of the system affects the signal propagation inside a waveguide, and hence, the propagation model must capture the frequency-dependent effects. 

The microstrip line comprises a conductor of width  $\upsilon$, printed on a dielectric substrate with thickness $\zeta$ and dielectric constant $\epsilon_r$. Thus, the effective dielectric constant at DC is~\cite{pozar2011microwave}
\begin{equation}\label{eq:effdielecDC}
    \epsilon'_{e} = {(\epsilon_r+1)}/{2} + {(\epsilon_r-1)}/{2}{\sqrt{1 + 12{\zeta}/{\upsilon}}}.
\end{equation}
The effective dielectric constant at frequency $f$ is~\cite{pozar2011microwave}
\begin{equation}
    \epsilon_e^f = \epsilon_r - {\bigl(\epsilon_r-\epsilon'_e\bigr)}/{\bigl(1 + G(f)\bigr)},
\end{equation}
where $G(f) = {(0.6 + 0.009Z_0)f^2}/{ (Z_0/8\pi \zeta)^2}$, and
\begin{equation} \label{eq:Z0}
    Z_0 = \begin{cases} 60 \ln{(8\zeta/\upsilon + \upsilon/(4\zeta))}/\sqrt{\epsilon^f_e}, & \upsilon \leq \zeta, \\ \frac{120\pi/\sqrt{\epsilon^f_e}}{\upsilon/\zeta + 1.393+0.667\ln{(\upsilon/\zeta + 1.444)}}, &  \upsilon \geq \zeta, \end{cases}
\end{equation} 
is the characteristic impedance of the microstrip, while both cases in \eqref{eq:Z0} are approximately equal for ${\upsilon}/{\zeta} = 1$. Thereby, $\beta = ({2\pi}/{\lambda})\sqrt{\epsilon^f_e}$
is the microstrip propagation constant and the attenuation due to the dielectric loss is given by 
\begin{equation}
    \alpha_d = \pi\epsilon_r(\epsilon^f_e - 1)\varrho/\Bigl({\lambda\sqrt{\epsilon_e^f}(\epsilon_r -1)}\Bigr),
\end{equation}where $\varrho$ is the loss tangent of the dielectric. Moreover, $\alpha_c = R_s/Z_0\upsilon$ is the approximate attenuation due to the conductor loss and $R_s = \sqrt{2\pi f \mu_0/2\sigma}$ is the conductor surface resistivity with $\sigma$ and $\mu_0$ being the conductivity and the free space permeability, respectively. Finally, the microstrip line attenuation coefficient is modeled as $\alpha = \alpha_d + \alpha_c$. Table~\ref{tab:material} lists some microstrip materials with their main characteristics. 

\renewcommand{\arraystretch}{1.1}
\begin{table}[t]
\centering
\caption{Microstrip material characteristics.}
\vspace{-2mm}
    \begin{tabular}{@{} |l|l|l|l|}
    \hline 
        \textbf{Material} & $\mathbf{\boldsymbol{\epsilon}_r}$ & $\boldsymbol{\varrho}$ &  \textbf{Typical $\boldsymbol{\zeta}$\protect\footnotemark} \\
    \hline
        \textregistered Cylex FR4 @50Hz & 5.5 & 0.04  &  1.6 mm  \\
        \textregistered DuPont Pyralux AP-9161 @1MHz& 3.4   & 0.002  &  0.15 mm \\
        \textregistered Arlon AD260A @1 MHz & 2.6 & 0.00135 &  1.14 mm  \\
        \textregistered Rogers RO4003C @2.5GHz & 3.55  & 0.0021 &  0.53 mm  \\
        \textregistered Taconic RF-10 @10GHz & 10.2 & 0.0025 &  0.25 mm  \\
        \textregistered Panasonic R-5515 @14, 26GHz & 3 & 0.002 &  0.105 mm  \\
        \textregistered Isola IS580G @5, 10, 20GHz & 3.8 & 0.006 &  0.4 mm  \\
    \hline
    \end{tabular}

    \label{tab:material}
\end{table}

\footnotetext{The thickness of the material is not strictly limited to the mentioned value and the typical $\zeta$ serves as an indicator of the product thickness range.} 

Next, $\mathbf{H} \in \mathbb{C}^{N \times N}$ is the microstrip propagation diagonal matrix with $h_{i,l}$ being the $\bigl((i-1)N_h + l\bigr)$th column and $\bigl((i-1)N_h + l\bigr)$th row element. The Lorentzian-constrained phase model is utilized to capture the dependency between the elements amplitude and phase \cite{DMAformulation}, so that the frequency response of the $l$th element in the $i$th microstrip is 
\begin{equation}\label{eq:lorentzweight}
    q_{i, l} \in \mathcal{Q} = \Big\{{ (j+e^{j \phi_{i,l}})}/{2} \Big| \phi_{i,l} \in[0,2 \pi]\Big\}, \quad \forall i, l.
\end{equation}
Additionally, $\mathbf{Q} \in \mathbb{C}^{N \times N_d}$ is the matrix containing the configurable weights of the metamaterial elements \cite{near-field}, i.e.,
\begin{equation} \label{eq:Q}
    \mathbf{Q}_{(i-1) N_h+l, n}= \begin{cases}q_{i, l}, & i=n, \\ 0, & i \neq n .\end{cases}
\end{equation}

\subsubsection{Signal Model}\label{subsec:signal}

Let $x_m$ be the $m$th energy symbol at the input of the digital precoder where $m = 1,\ldots, M$ and $M = \min(N_v, K)$. Also, the energy symbols are independent and normalized such that $\mathbb{E}\bigl\{x_n^H x_r\bigr\} = 0$ and $\mathbb{E}\bigl\{x_n^H x_n\bigr\} = 1$, while $\mathbf{w}_m$ is the $N_v$-dimensional precoding vector for $x_m$. Thus, the transmit signal is $\mathbf{s} = \sum_{m = 1}^{M} \mathbf{H Q w}_mx_m$, while the transmit power is
        $P_{Tx} = \mathbb{E}_x\bigl\{\mathbf{s}^H \mathbf{s}\bigr\} 
    = \sum_{m = 1}^{M} |\mathbf{H Q w}_m|^2$.
The received energy signal at the $k$th EH device is given by $y_k = \boldsymbol{\gamma}_k^H\mathbf{s}$, while the corresponding received RF power is 
\begin{equation}\label{eq:rxpower}
    P_{Rx}^k = \mathbb{E}_x\bigl\{y_k^H y_k\bigr\} = \sum_{m = 1}^{M} |\boldsymbol{\gamma}_k^H \mathbf{H Q w}_m|^2.
\end{equation}

\vspace{-2mm}
\subsection{Problem Formulation}\label{subsec:probform}

In practical WPT systems, the transmit power influences largely the total power consumption. Indeed, the transmit power determines the power consumption of the PAs, which are the most power-hungry system components. Motivated by this, our objective is to minimize the transmit power while satisfying the users' RF power requirements. Assuming the location of the users is known, the optimization problem is
\begin{subequations}\label{prob1}
\begin{align}
\label{prob1a} \mathrm{\textbf{\textsc{P1}}:} \quad \minimize_{\mathbf{Q}, \{\mathbf{w}_m\},\forall m} \quad & \sum_{m = 1}^{M} |\mathbf{H Q w}_m|^2\\
\textrm{subject to} \label{prob1b} \quad & \sum_{m = 1}^{M} |\boldsymbol{\gamma}_k^H \mathbf{H Q w}_m|^2  \geq \delta_k , \quad \forall k,\\
&  \label{prob1c} (\ref{eq:Q}),\quad q_{i,l} \in \mathcal{Q}, \quad \forall i,l.
\end{align}
\end{subequations}
where $\delta_k$ is the RF power meeting the EH requirements of the $k$th user. The problem is non-convex while the variables are highly coupled, due to the correlation between the phase and amplitude of the metamaterial elements through the Lorentzian constraint, making (\ref{prob1}) very difficult to be optimally solved.

\vspace{-2mm}
\section{Proposed EB Optimization Solution}\label{sec:optframework}

\subsection{Optimal Digital Precoders with Fixed $\mathbf{Q}$}

Let us fix $\mathbf{Q}$ and rewrite (\ref{eq:rxpower}) as
\begin{flalign}\label{eq:reformQfix1}
    P_{Rx}^k &= \sum_{m = 1}^M\bigl((\boldsymbol{\gamma}_k^H \mathbf{H Q}) \mathbf{w}_m\bigr)^H (\boldsymbol{\gamma}_k^H \mathbf{H Q}) \mathbf{w}_m  \nonumber\\ 
    &= \mathrm{Tr}\biggl(\biggl[\sum_{m = 1}^M \mathbf{w}_m \mathbf{w}_m^H\biggr] \mathbf{b}_k \mathbf{b}_k^H  \biggr) =
    \mathrm{Tr}\bigl(\mathbf{W}\mathbf{B}_k\bigr),
\end{flalign}
which comes from  several algebraic transformations and by defining $\mathbf{b}_k = (\boldsymbol{\gamma}_k^H \mathbf{HQ})^H \in \mathbb{C}^{N_d \times 1}$, $\mathbf{W} = \sum_{m = 1}^M \mathbf{w}_m\mathbf{w}_m^H$, and $\mathbf{B}_k = \mathbf{b}_k \mathbf{b}_k^H$. Similarly, we reformulate the transmit power as
    $P_{Tx} = 
    \mathrm{Tr}\bigl(\mathbf{W}\mathbf{F}\bigr)$,
where $\mathbf{F} = (\mathbf{HQ})^H \mathbf{HQ}$. By leveraging (\ref{eq:reformQfix1}), (\ref{prob1}) can be reformulated with a fixed $\mathbf{Q}$ as 
\begin{subequations}\label{probfixedQ}
\begin{align}
 \label{probfixedQ1} \mathrm{\textbf{\textsc{P2}}:} \quad \minimize_{\mathbf{W}} \quad & \mathrm{Tr}\bigl(\mathbf{W}\mathbf{F}\bigr)\\
\textrm{subject to} \label{probfixedQ2} \quad & \mathrm{Tr}\bigl(\mathbf{W}\mathbf{B}_k\bigr) \geq \delta_k, \quad \forall k,\\
&  \label{probfixedQ3} \mathbf{W} \succeq \mathbf{0},
\end{align}
\end{subequations}
which is an SDP that can be solved by standard convex optimization tools, e.g., CVX \cite{cvxref}. Then, the optimal precoding vectors $\{\mathbf{w}_m\}_{\forall m}$ can be obtained as the eigenvectors of $\mathbf{W}$ multiplied by the square root of their respective eigenvalues. 
\vspace{-2mm}
\subsection{Suboptimal $\mathbf{Q}$ with Fixed Digital Precoders}

Now, let us fix $\{\mathbf{w}_m\}_{\forall m}$. Then, utilizing the fact that $\mathbf{a}^T \mathbf{G} \mathbf{b} = (\mathbf{b}^T \otimes \mathbf{a}^T) \mathrm{Vec}(\mathbf{G})$ , we obtain \cite{near-field}
\begin{align}
\label{eq:matforQ1} |\boldsymbol{\gamma}_k^H \mathbf{H Q w}_m|^2 &= |(\mathbf{w}_m^T \otimes 
(\boldsymbol{\gamma}_k^H \mathbf{H})) \mathbf{q}|^2 = |\mathbf{z}_{m,k}^H \mathbf{q}|^2,
\end{align}
where $\mathbf{q} = \mathrm{Vec}(\mathbf{Q})\in \mathbb{C}^{N N_v \times 1}$ and $\mathbf{z}_{m,k} = (\mathbf{w}_m^T \otimes 
(\boldsymbol{\gamma}_k^H \mathbf{H}))^H \in \mathbb{C}^{N N_v \times 1}$. Moreover, we define $\hat{\mathbf{q}} \in \mathbb{C}^{N\times1}$ as the modified version of $\mathbf{q}$ without the zero elements, thus, $\hat{\mathbf{q}}=[q_{1,1}, q_{1,2}, \cdots,q_{N_v,N_h}]^T$. Additionally, $\hat{\mathbf{z}}_{m,k} \in \mathbb{C}^{N\times1}$ is obtained by removing the elements with the same index as the zero elements in $\mathbf{q}$. Then, the RF power received at user $k$ is  
\begin{flalign}\label{eq:reformWfix1}
    P_{Rx}^k  &\labelrel={reformWfix1:1} \sum_{m = 1}^M (\mathbf{z}_{m,k}^H \mathbf{q})^H (\mathbf{z}_{m,k} \mathbf{q}) \nonumber \\
     &\labelrel={reformWfix1:2} \hat{\mathbf{q}}^H \biggl[\sum_{m = 1}^M \hat{\mathbf{z}}_{m,k} \hat{\mathbf{z}}_{m,k}^H \biggr] \hat{\mathbf{q}} 
     \labelrel={reformWfix1:3} \mathrm{Tr}\bigl(\widetilde{\mathbf{Z}}_{m,k} \widetilde{\mathbf{Q}}\bigr),
\end{flalign}
where \eqref{reformWfix1:1} comes from using (\ref{eq:matforQ1}) followed by a simple transformation and \eqref{reformWfix1:2} from zeroing the terms impacted by the zero elements of $\mathbf{q}$. Furthermore, \eqref{reformWfix1:3} comes from defining $\widetilde{\mathbf{Q}} = \hat{\mathbf{q}}\hat{\mathbf{q}}^H$ and $\widetilde{\mathbf{Z}}_{m,k} = \sum_{m = 1}^M \hat{\mathbf{z}}_{m,k} \hat{\mathbf{z}}_{m,k}^H$. As previously mentioned, the Lorentzian-constrained phase of the elements makes the problem very complex and difficult to solve. One way to cope with this complexity is by utilizing approximations to relax \eqref{prob1c} but this may cause the solution to violate the constraint \eqref{prob1b}. Instead, we propose decoupling the problem to first maximize the minimum RF power at the users given a fixed $\mathbf{W}$ and leveraging only the beamforming capability of the DMA elements, and then utilizing such a feasible solution to determine the $\mathbf{W}$ that minimizes the transmit power consumption subject to \eqref{prob1b} and given fixed $\widetilde{\mathbf{Q}}$. By conducting this procedure iteratively, a suboptimal solution is obtained. Notice that such an optimization approach adapts well to the fact that the gain introduced by the metamaterial elements is limited and correlated with their phase values, while the gain in digital precoders can be chosen freely without any limitation and correlation with the phase values, making them the dominant factor in determining the transmit power. Thereby, the optimization problem is formulated as
\begin{subequations}\label{probfixedW}
\begin{align}
\label{probfixedW1} \mathrm{\textbf{\textsc{P3}}:} \quad  \maximize_{\widetilde{\mathbf{Q}}} \quad &  \min_k \mathrm{Tr}\bigl(\widetilde{\mathbf{Z}}_{m,k} \widetilde{\mathbf{Q}}\bigr)\\
\textrm{subject to} \quad &\label{probfixedW3} \widetilde{\mathbf{Q}} \succeq \mathbf{0},\\
&\label{probfixedW4} q_{i,l} \in \mathcal{Q}, \quad \forall i,l, \\
&\label{probfixedW5} \mathrm{rank}(\widetilde{\mathbf{Q}}) = 1.
\end{align}
\end{subequations}

\begin{figure}[t]
    \centering
    \includegraphics[width=0.5\columnwidth]{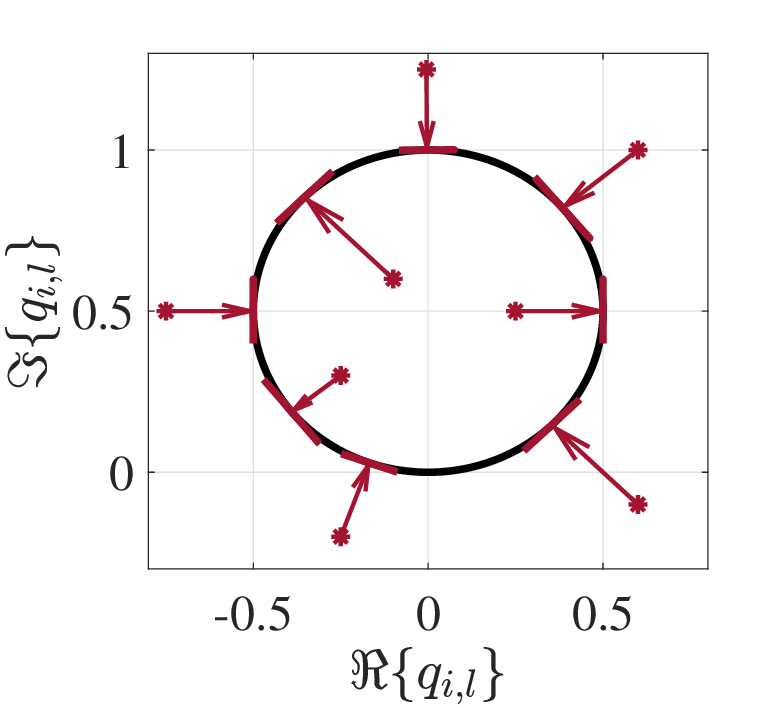}
    \caption{Mapping the $\mathrm{\textbf{\textsc{P4}}}$ solution to the Lorentzian constrained weights in the complex plane.}
    \label{fig:phasemodel}
    \vspace{-4mm}
\end{figure}

Problem (\ref{probfixedW}) is still difficult to solve due to the Lorentzian and rank constraints. Therefore, we relax the problem as
\begin{subequations}\label{probfixedWrelax}
\begin{align}
\label{probfixedWrelax1} \mathrm{\textbf{\textsc{P4}}:} \quad \maximize_{\widetilde{\mathbf{Q}}, t} \quad &  t\\
\textrm{subject to} \label{probfixedWrelax2} \quad & t \leq \mathrm{Tr}\bigl(\widetilde{\mathbf{Z}}_{m,k} \widetilde{\mathbf{Q}}\bigr), \quad \forall k,\\
&\label{probfixedWrelax4} \mathrm{Tr}\bigl( \widetilde{\mathbf{Q}}\bigr) \leq N, \\
&\label{probfixedWrelax3} \widetilde{\mathbf{Q}} \succeq \mathbf{0},
\end{align}
\end{subequations}
where \eqref{probfixedWrelax4} prevents the problem from becoming unbounded and comes from the fact that the maximum value of the diagonal elements of $\widetilde{\mathbf{Q}}$ is $1$. Hence, the problem becomes an SDP that can be solved similarly to (\ref{probfixedQ}). Then, we transform the relaxed solution into a feasible solution according to constraints (\ref{probfixedW4}) and (\ref{probfixedW5}). First, $\mathbf{q}'$ is obtained as the dominant eigenvector of $\widetilde{\mathbf{Q}}$ multiplied by the square root of its eigenvalue. Still, $\mathbf{q}'$ is not Lorentzian constrained, thus, we map each of the elements in $\mathbf{q}'$ to the nearest point on the Lorentzian constrained circle as represented in Fig. \ref{fig:phasemodel}. Hereby, the configured phase of the $l$th metamaterial element in the $i$th microstrip can be obtained as
\begin{equation}\label{problorenz}
\phi^\star_{i,l} = \min_{\phi_{i,l}} \quad \Big|({j + e^{j{\phi}_{i,l}}})/{2} - q'_{i,l}\Big| , \quad \forall i,l,
\end{equation}
where $q'_{i,l}$ is the $\bigl((i-1)N_h + l\bigr)$th element in $\mathbf{q}'$, while (\ref{problorenz}) can be easily solved  using a one-dimensional search. Then, $q_{i,l}^\star = {(j + e^{j{\phi}^\star_{i,l}})}/{2}, \quad \forall i,l$.

 \vspace{-4mm}

\subsection{Overall Algorithm}

Algorithm~\ref{EBDMA_psedu_FD} summarizes the overall EB optimization solution for DMA-assisted RF-WPT system. At first, a random Lorentzian constrained frequency response is generated for each metamaterial element, and $\mathbf{Q}$ is constructed. Then, a local optimum solution is obtained for $\mathrm{\textbf{\textsc{P3}}}$ through lines \ref{alg1:line:P3_start}-\ref{alg1:line:P3_finish}, followed by finding the optimal precoders by solving $\mathrm{\textbf{\textsc{P2}}}$. Moreover, the solution is updated in lines \ref{alg1:line:up_start}-\ref{alg1:line:up_finish} if the transmit power decreases. The above process is repeated iteratively until there is no improvement in the solution for a maximum stall counter or a maximum number of iterations is reached.

The number of variables in \eqref{probfixedQ} and \eqref{probfixedWrelax} is ${N_v(N_v + 1)}/2$ and $1 + N(N + 1)/2$, respectively. Notice that the rest of the entries in $\mathbf{W}$ and $\widetilde{\mathbf{Q}}$ are determined according to the Hermitian structure, and the sizes of these Hermitian matrix sub-spaces are ${N_v}^2$ and $N^2$, respectively. Additionally, the number of constraints scales with $K$ in both problems. It is shown that for a given accuracy, the complexity of SDP problems grows at most with $\mathcal{O}(n^{1/2})$, where $n$ is the problem size, which is determined by the number of constraints and variables \cite{semidef-boyd}. Thus, the proposed algorithm, which solves two SDP problems for $I$ iterations in the worst case, converges in polynomial time and its complexity increases with $K$ and $N^2$.

\begin{algorithm}[t]
	\caption{EB Optimization for DMA-based RF-WPT.} \label{EBDMA_psedu_FD}
	\begin{algorithmic}[1]
            \State \textbf{Input:} $C$, $I$, $\{\boldsymbol{\gamma}_k, \delta_k\}_{\forall k}$ 
            \textbf{Output:} $\{\mathbf{w}^\star_m\}_{\forall m}$, $\{q^\star_{i,l}\}_{\forall i,l}$
            \State \textbf{Initialize:} \label{alg1:line:init_start}
            \State \hspace{6pt} \parbox[t]{\dimexpr\linewidth-\algorithmicindent}{Construct $\mathbf{Q}$ using random $\{\phi_{i,l}\}_{\forall i,l}$, (\ref{eq:lorentzweight}), and (\ref{eq:Q})}
            \State \hspace{6pt} Solve $\mathrm{\textbf{\textsc{P2}}}$ to obtain $P^\star_{Tx}$ and $\{\mathbf{w}^\star_m\}_{\forall m}$, $i' = 1$, $c' = 0$
            \Repeat
                \State \hspace{-2mm} Solve $\mathrm{\textbf{\textsc{P4}}}$ to obtain $\mathbf{q}'$ \label{alg1:line:P3_start}
                \State \hspace{-2mm} Obtain $\{q_{i,l}\}_{\forall i,l}$ using (\ref{problorenz}) and construct $\mathbf{Q}$\label{alg1:line:P3_finish}
                \State \hspace{-2mm} Solve $\mathrm{\textbf{\textsc{P2}}}$ to obtain $P_{Tx}$ and $\{\mathbf{w}_m\}_{\forall m}$, 
                \State \hspace{-2mm} $c' \leftarrow c' + 1$,\ $i' \leftarrow i' + 1$
                \label{alg1:line:init_start}
                \If{$P_{Tx} < P_{Tx}^\star$}\label{alg1:line:up_start}
                    \State \hspace{-4mm} $P^\star_{Tx} \leftarrow P_{Tx}$, $\mathbf{w}^\star_m \leftarrow \hspace{-2mm}\mathbf{w}_m$, $q_{i,l}^\star \leftarrow q_{i,l}$, $\forall {m, i,l}$, \ $c' \leftarrow 0$ 
                \EndIf\label{alg1:line:up_finish}
            \Until{$c' = C$ or $i' = I$}
\end{algorithmic} 
\end{algorithm}

\vspace{-2mm}
\section{Numerical Results}\label{sec:results}

We consider a 100 m$^2$ indoor area with a transmitter at the center of the ceiling with a 3 m height. The users are randomly distributed over 30 realizations and $\delta_k = 100$ $\mu$W, $\forall k$, while the spacing between the DMA metamaterial elements and the microstrips are $\lambda/5$ and $\lambda/2$, respectively. Thus, $N_v = \lfloor \frac{L}{\lambda/2} \rfloor$ and $N_h = \lfloor \frac{L}{\lambda/5} \rfloor$. A \textregistered DuPont Pyralux AP-9161 is utilized to design the microstrip line while its propagation coefficients are estimated for different frequencies as in Section~\ref{subsec:DMAcharach}.

We refer to the proposed method as EB-ASD, while particle swarm optimization (PSO) \cite{PSOmain} with 1000 iterations and 100 particles per iteration is used as a benchmark. The DMA performance is compared to the FD structure with an inter-element distance of $\lambda/2$. The optimal FD precoders are obtained by solving $\mathrm{\textbf{\textsc{P2}}}$ with $\mathbf{b}_k = \boldsymbol{\gamma}_k$ as in \cite{OnelRadioStripes}. Notably, an FD setup requires a considerably large number of RF chains with high power consumption, which is not considered here. Thus, the performance gap between DMA and FD may be even larger in practice. Assuming $U$ random locations for a device, the average gain in the figures is $\mathbb{E}_u\bigl\{|\boldsymbol{\gamma}_u|^2\bigr\}$ and $\mathbb{E}_u\bigl\{|\boldsymbol{\gamma}_u^H \mathbf{H}|^2\bigr\}$ with $u = 1, \ldots, U$ for FD and DMA, respectively.


\begin{figure}[t]
\centering
    \begin{subfigure}[b]{0.85\columnwidth}
        \includegraphics[width=\columnwidth]{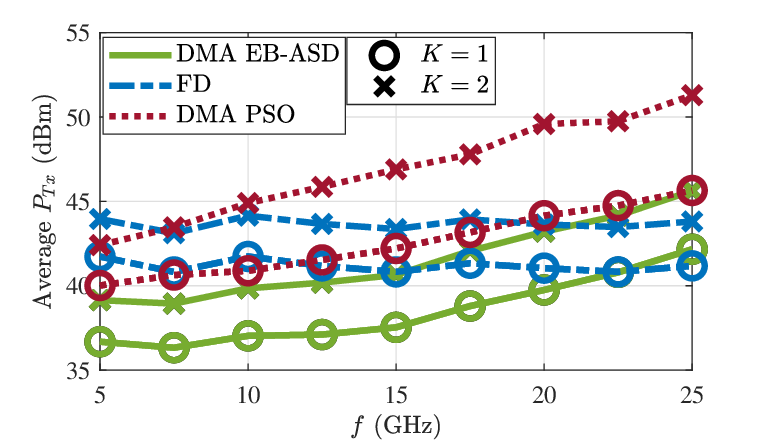}
        \label{fig:overF_P}
        \hfill
        \includegraphics[width=\columnwidth]{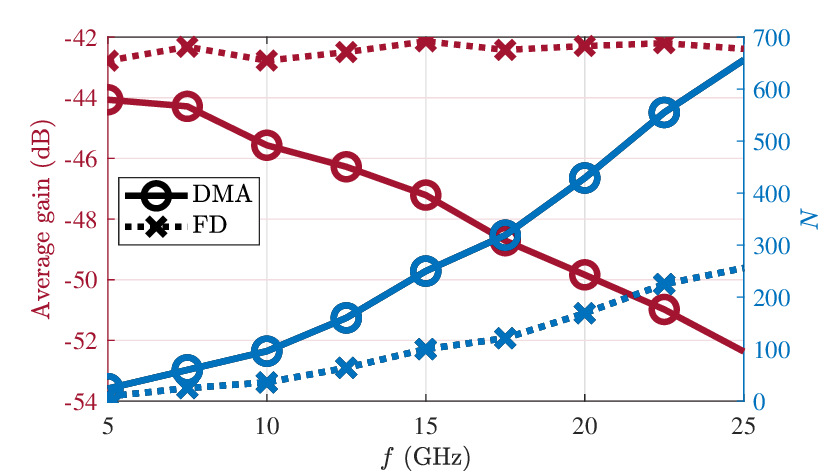}
        \label{fig:overF_Loss}
    \end{subfigure}
    \vspace{-3mm}
    \caption{(a) Average transmit power (top) and (b) average gain and the number of antenna elements (bottom), as a function of frequency for fixed $L = 10$ cm, $K = 1$, and $K = 2$.}
    \label{fig:overF}
    \vspace{-3mm}
\end{figure}

Fig.~\ref{fig:overF} illustrates the simulation results over different operating frequencies. It is observed that the proposed approach outperforms PSO. The reason is that the problem space is large, especially in high-frequency regime, and thus, PSO needs much more time and particles to perform well. For a multi-user setup, maximum ratio transmission (MRT)-based \cite{OnelRadioStripes} precoders can be utilized to derive a lower bound for the transmit power of the FD structure. Hereby, one may expect that the received RF power scales with $P_{Tx, k}|\boldsymbol{\gamma}_k|^2$, where $\sum_{m = 1}^M P_{Tx, m} = P_{Tx}$. Moreover, the transmit power should be divided among $K$ users and meet all RF power requirements, thus, the transmit power increases with $K$, while it slightly changes over $f$ since $|\boldsymbol{\gamma}_k|^2$ almost remains fixed. In the DMA, one may expect that the received RF power by the $k$th user is influenced by the pattern of $|\boldsymbol{\gamma}_k^H \mathbf{H}|^2$ since it includes the majority of the losses. However, there is no guarantee that the pattern is just influenced by this term and the structure of $\mathbf{Q}$ and its values are also important. Therefore, although $|\boldsymbol{\gamma}_k^H \mathbf{H}|^2$ decreases, the increasing pattern of the transmit power with frequency is not assured and one may experience other behaviors in different frequency ranges, as seen in the low-frequency regime in the figure. All in all, the DMA transmit power pattern may change depending on the scenario and its parameters, such as the antenna form factor, distance to the users, and the number of users. Similar to FD, the transmit power increases with $K$ since the total power should be split among more users to meet their requirements. Meanwhile, the number of elements is larger in DMA leading to more beamforming capability, and additionally, one signal can feed multiple elements in the DMA. Thus, the total output power of the RF chains (transmit power) is less compared to FD, and DMA outperforms FD in the low-frequency regime in terms of the required transmit power. However, the DMA losses may become large in higher frequencies depending on the scenario and antenna form factor, and FD may perform better.

\begin{figure}[t]
\centering
    \begin{subfigure}[b]{0.85\columnwidth}
        \includegraphics[width=\columnwidth]{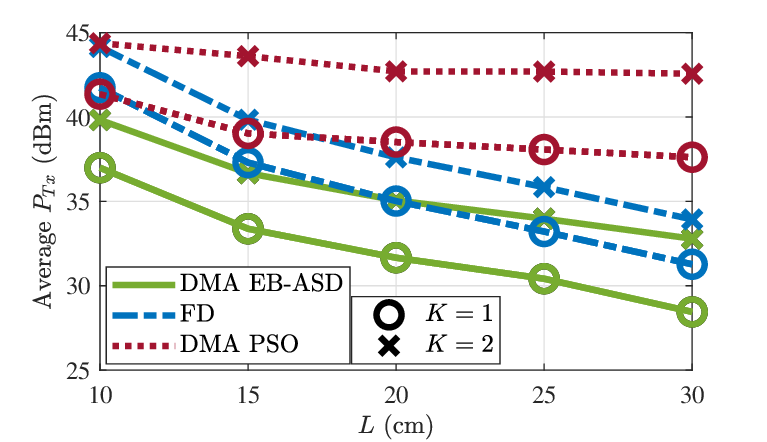}
        \label{fig:overL_P}
        \hfill
        \includegraphics[width=\columnwidth]{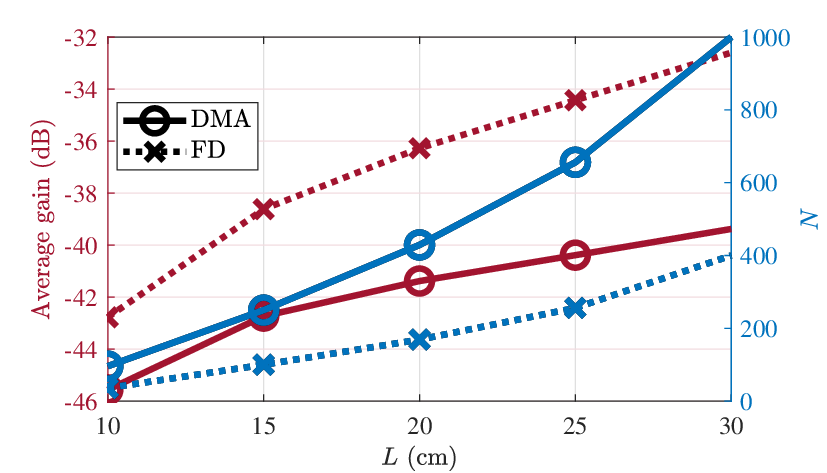}
        \label{fig:overL_Loss}
    \end{subfigure}
    \vspace{-4mm}
    \caption{(a) Average transmit power (top) and (b) average gain and the number of antenna elements (bottom), as a function of antenna length for fixed $f = 10$ GHz, $K = 1$, and $K = 2$.}
    \label{fig:overL}
    \vspace{-3mm}
\end{figure}

The influence of the antenna length is illustrated in Fig.~\ref{fig:overL}. The discussions of the previous case are also applicable here. As expected, the transmit power decreases with antenna length in both architectures since the average effective gain increases. Additionally, DMA outperforms FD in this specific frequency but the situation may change at significantly higher frequencies, as DMA losses become considerably large. The proposed approach also outperforms PSO in this configuration.

\vspace{-2mm}
\section{Conclusion}\label{sec:consclusion}

We considered a multi-user RF-WPT system with a DMA-based architecture. Moreover, we proposed an EB design aiming to minimize the transmit power while meeting the users' EH requirements. The solution, using SDP and alternating optimization, outperformed PSO, and our results also reveal that a DMA transmitter is preferred to the FD alternative. It was observed that although increasing the system frequency does not affect the FD architecture performance, it increases the required transmit power for the DMA-assisted system since the effective gain decreases. Meanwhile, increasing the number of elements by utilizing a larger antenna array may lead to significant performance gains in both architectures.


\vspace{-2mm}
\bibliographystyle{ieeetr}
\bibliography{ref_manuall_abb}


%




\end{document}